# Sheath effects with thermal electrons on the resonance frequency of a DC-biased hairpin probe

Pawandeep Singh[1,2,(a)], Avnish Pandey[3], Swati Dahiya[1,2], Yashashri Patil[1], Nishant Sirse[4], Shantanu Karkari[1,2]

[1]Institute for Plasma Research, Bhat, Gandhinagar, Gujarat, 382428 India

[2]Homi Bhabha National Institute, Training School Complex, Anushaktinagar, Mumbai-400094, India

[3]Department of Applied Science and Humanities, United College of Engineering and Research, Prayagraj—211009, India

[4]Institute of Science and Research and Centre for Scientific and Applied Research, IPS Academy, Indore-452012, India

[a)] corresponding author: singh.pawandeep67@gmail.com

___

## Abstract

The dielectric constant of a sheath, whether ionic or electronic, formed around the cylindrical limbs of a hairpin probe, is often considered the same as that of a vacuum. However, this assumption does not hold true for electron sheaths and electron-permeating ionic sheaths, resulting in a deviation of the sheath dielectric constant from that of a vacuum. This deviation significantly influences the effective dielectric between the cylindrical limbs. As a result, it impacts the theoretically estimated resonance frequency characteristic curve of a DC-biased hairpin probe. In this study, we investigate the influence of electron temperature on the sheath dielectric and, consequently, on the resonance frequency characteristic curve. The findings shows that electron temperature primarily determines the resonance frequency characteristic curve. With increasing electron temperature, the peak in the resonance frequency characteristic curve shifts towards higher positive probe bias values and exhibits a broadening near the maxima instead of a sharp peak. This broadening near the maxima has also been validated with an experimentally measured resonance frequency characteristic curve in a capacitively coupled argon discharge.

___

## 1. Introduction

Hairpin probe (HP) is a device to measure the absolute electron density in a discharge [1–6]. Because of its direct and reproducible results, HP has been adopted in various plasma systems, such as reactive plasmas [7–9], pulse discharges [10–12], and laser photodetachment diagnostics [13] for phase- and time-resolved measurements [14,15]. More recently, the applicability of the HP has been extended to measure electron temperature and sheath width as a function of applied DC bias [16], negative ion density, and temperature by applying a train of negative voltage pulses [17,18].

HP is based on the resonance detection principle of a quarter-wave parallel wire transmission line [1,6], in which the resonance frequency ($f$) solely depends on the length ($L$)/width ($w$) of the HP and the effective permittivity ($k_{eff}$) of the medium, as

$$f_r = \frac{c/2(2L+w)}{\sqrt{k_{eff}}} \qquad (1)$$

In vacuum, $k_{eff} \to 1$; hence the vacuum resonance frequency becomes, $f_0 = c/2(2L+w)$. However, in plasma, the effective permittivity consists of permittivity of the cylindrical sheath around the HP' limbs and permittivity of the plasma. In low-pressure, un-magnetized collision-less plasma [19], under cold ion approximation, the plasma permittivity ($k_p$) directly relates to the electron plasma frequency ($f_{pe} = \sqrt{n_e e^2/m_e \varepsilon_0}$) and hence to absolute value of electron density ($n_e$) as

$$k_p = 1 - \frac{n_e e^2/m_e \varepsilon_0}{f_r^2}. \qquad (2)$$

Whereas, the sheath dielectric ($k_d$) is generally assumed to be vacuum dielectric, i.e., $k_d = 1$, while the sheath radius '$b$' depends on the applied bias [6,20–23]. Therefore, sheath can be assumed to be merely a geometrical effect altering the effective permittivity ($k_{eff}$) and hence electron density ($n_e$) measurement using HP.

To determine this geometrical effect on the effective permittivity ($k_{eff}$), Piejak et al. [24] have modelled sheath and plasma as capacitors in series. The values for the respective capacitances per unit length are derived using method of images reducing the problem to an infinite wire of radius '$a$' and medial-plane from the two infinitely long parallel wires having separation width '$2h$' and derived a sheath correction factor ($\xi$) for the resonance frequency as

$$f_{pe}^2 = \frac{f_r^2 - f_0^2}{\xi} \quad (3a)$$

where $\xi = 1 - \frac{f_0^2}{f_r^2}\Lambda,$ (3b)

and $\Lambda = 1 - \ln\left(\frac{2h-b}{b}\right)/\ln\left(\frac{2h-a}{a}\right).$ (3c)

However, the sheath correction factor ($\xi$) obtained by Piejak et al. and later on by others [20,21,23,24] relies on the assumption that the sheath dielectric is close to the vacuum dielectric constant value, leading to misapprehension that the ionic sheath has no electrons. Though this is partly true, when the DC probe is biased at large negative value compared to the plasma potential. When the applied bias is close to plasma potential, energetic electrons can adequately enter the sheath and will change the sheath dielectric. Considering this fact, Haas et al. [25] have evaluated the sheath dielectric constant ($k_d$) using the cold plasma permittivity [Eq. (2)] wherein the electron density is replaced with average electron density ($\bar{n}_e$) integrated over the sheath, for both ion and electron sheath in an electropositive plasma. Consequently, a biquadratic equation is derived for the resonance frequency ($f$) using the capacitive model, as

$$k_d f_r^4 - \left(k_d(1-\Lambda)f_0^2 + \Lambda f_0^2 + k_d f_{pe}^2\right)f_r^2 + \Lambda f_0^2 f_{pe}^2 = 0 \quad (4)$$

However, in the formulation, the electron density ($n_{es}$) at the sheath edge was considered to be the same as bulk ($n_e$), i.e., edge-to-centre density fall was ignored. The present study revisits Haas's model, incorporating the edge-to-centre density fall by relating it to the presheath potential fall and determining the influence of electron temperature on the resonance frequency characteristics. Subsequently, the resonance frequency characteristic curve, obtained from this model, is compared with the experimentally derived curve in an annular parallel plate capacitively coupled plasma discharge. This paper has been organised as follows:

In Section 2, a brief description of the electron and ion sheath model is presented, whereas its impact on the resonance frequency is presented in Section 3. The specifics of the experimental setup and diagnostics are provided in Section 4, while the experimental findings are presented in Section 5. A brief discussion with a summary highlighting the key findings of the work is presented in Section 6.

## 2. Electron and ion sheath around DC biased Hairpin resonator probe

When the HP is biased with a DC potential ($V_B$) with respect to plasma potential ($V_P$), an ionic sheath ($V_B < V_p$) or electronic sheath ($V_B > V_p$) of cylindrical shape is formed around both of its limbs. In either of the case, the resonance frequency of the HP [Eq. (4)] will depend upon the sheath radius ($b$) [via $\Lambda$ in Eq. (3b)] as well as its dielectric ($k_d$). The following subsections briefly presents the theoretical model for ion and electron sheath to estimate the value of $b$ and $k_d$.

### 2.(a) Determining sheath radius (b):

*Ion Sheath:*

To determine the value of sheath radius ($b$) as a function of bias ($V_B$), the sheath potential profile, $\varphi(r) = V_S - V_b(r)$, is solved by using the Poisson's equation in cylindrical coordinates [Eq. (5)], where $V_S$ is potential at the sheath edge.

$$\frac{d^2\varphi}{dr^2} + \frac{1}{r}\frac{d\varphi}{dr} + \frac{en_{es}}{\varepsilon_0}\left[\frac{b}{r}\frac{1}{\sqrt{\left(1-\frac{2e\varphi}{Mu_B^2}\right)}} - \exp\left(\frac{\varphi}{T_e}\right)\right] = 0 \quad (5)$$

In the above equation, $n_{es}$ is electron density at the sheath edge, while $e$ is the electronic charge, $\varepsilon_0$ is absolute permittivity of free space, $M$ is positive ion mass, $T_e$ (in eV) is electron temperature, and $u_B = \sqrt{eT_e/M}$ is the Bohm speed [26].

The third term in Eq. (5) represents the total space charge. In which, the positive ion density is represented in terms of potential using the continuity and momentum equation for positive ions in a collision-less sheath while considering the ions to be falling radially inwards onto the probe surface [16]. Whereas, electrons are represented as Boltzmann distribution.

For a given plasma parameters, $T_e$ and $n_e$, Eq. (5) can be solved for $\varphi(r)$ as a boundary-value-problem. Since, sheath radius ($b$) is not known a-priori, the Eq. (5) is generally solved as an initial-value-problem (IVP) by supplying a value for the potential, $\varphi(b)$ and electric field at the sheath edge, $\varphi'(b)$. Without the loss of generality, the value of potential at the sheath edge is assumed to be $\varphi(b) \approx 0$. Whereas, the value of electric field at the sheath edge is taken as $\varphi'(b) \approx 0.962\, T_e/(e\lambda_{Ds})(2\lambda_{Ds}Z/c_s)^{3/5}$. This value of sheath edge electric field was proposed by Kaganovich et al [27] by solving the transition region in between the quasi-neutral presheath region and non-neutral ionic sheath region considering the ionization collisions. Subsequently, the voltage at the probe surface is determined corresponding to each assumed values of $b$ by examining $\varphi(a) = V_B$ condition. Here, $\lambda_{Ds}$ is the Debye length corresponding to electron density at the sheath edge and Z is the ionization frequency which is assumed to be constant in the theoretical analysis, $Z = 10^{-15} m^3/s$.

*Electron Sheath:*

For $V_B > V_P$, owing to the small area of HP, an electron sheath forms [28–30]. Similar to the ionic sheath, electrons within the electron sheath are accelerated towards the probe due to electric field, resulting in a rise in potential towards the probe. Hence, the Poisson equation for evaluating $\varphi(r) = V_P - V_b(r)$ reads;

$$\frac{\partial^2 \varphi}{\partial r^2} + \frac{1}{r}\frac{\partial \varphi}{\partial r} - \frac{en_{es}}{\varepsilon_0}\left[\frac{b}{r}\frac{1}{\sqrt{1+\frac{2e\varphi}{m_e v_{es}^2}}}\right] = 0 \qquad (6)$$

where $m_e$ and $v_{es}$ are mass and speed of electron at the sheath edge.

Since the positive ion temperature remains close to the background gas neutrals in low pressure plasma, the positive ion contribution in Eq. (6) is ignored. However, similar to the case of ions in the ionic sheath, electrons in the electron sheath are considered as falling radially onto the probe and the densities are represented in terms of potential using the continuity and momentum equations. The velocity of electrons at the sheath edge is assumed to be its respective thermal speed $v_{es} = v_{eth} \equiv \sqrt{8eT_e/\pi m_e}$, while density is assumed to be the bulk densities, $n_{es} = n_e$. Hence, for a given values of $n_e$, and $T_e$, Eq.-(6) can be solved as an IVP, $\varphi'(b) \approx 0$, and $\varphi(b) \approx 0$, to determine the sheath potential profile and sheath radius ($b$).

*2.(b) Determining sheath dielectric:*

In the ion sheath, the highly energetic electrons belonging to the tail of the Maxwellian distribution can enter the ionic sheath. These infiltrated electrons act the same as the Boltzmann electrons present in the bulk, but with reduced density towards the probe surface. The average electron density inside the sheath of radius ($b$) as a result of an applied bias of $\varphi_B \equiv V_P - V_B$ at the HP, can be written as

$$\bar{n}_e = n_{es}\frac{T_e}{|\varphi_B|}\left[1 - \exp\left(-\frac{|\varphi_B|}{T_e}\right)\right] \qquad (7)$$

Hence, sheath dielectric ($k_d$) can be evaluated using the cold plasma permittivity [Eq. (2)] wherein the electron density is replaced with average electron density ($\bar{n}_e$) integrated over the sheath [31], as

$$k_d = 1 - \frac{f_{pe}^2}{f^2}\cdot\frac{e^{-\eta_s}T_e}{|\varphi_B|}\{1 - \exp\left(-\frac{|\varphi_B|}{T_e}\right)\} \qquad (8)$$

wherein the electron density at the sheath edge is written in terms of bulk electron density as $n_{es} = n_e e^{-\eta_s}$ employing the Boltzmann relation. Here, $\eta_s = (V_P - V_S)/T_e$ is the normalised presheath potential fall and its value has been taken as $\eta_s = 0.693$, $V_S$ is the potential at the ion sheath edge.

Similarly, the permittivity for electron sheath can be obtained as follows:

$$k_d = 1 - \frac{f_{pe}^2}{f^2}\cdot\frac{1}{b-a}\int_a^b \frac{b}{r\sqrt{1+\frac{\pi\varphi}{4T_e}}}dr \qquad (9)$$

Therefore, the sheath dielectric constant for ion and electron sheath can be determined using Eq. (8) and (9). The following section described the theoretical results obtained for different values of plasma parameters.

## 3. Theoretical Results:

Fig. (1) shows the ratio of sheath radius to probe radius ($b/a$) as a function of applied bias. The electron temperature ($T_e$) is varied, and the ion and electron sheath are separately solved using Eqs. (5) and (6) and merged at $\varphi_B = 0$. Consequently, $\varphi_B < 0$ signifies the ion sheath, whereas $\varphi_B > 0$ signifies the electron sheath.

It can be seen that the sheath diminishes to zero at $\varphi_B = 0$, i.e., $b = a$, while it increases with increase in the applied bias, either positive or negative [Fig.-1]. For a fixed negative (or positive) bias, the sheath width reduces with rise in $T_e$. This is because, with the increase in electron temperature ($T_e$), Bohm velocity increases which is the synergistic effect of increase in the presheath potential fall ($\sim T_e$), resulting an increased ionic flux at the sheath edge. Meanwhile, the electron flux that enter into the sheath also increases with $T_e$ due to rise in their mean energy. The increased flux of ions and electrons towards the probe would change the potential fall across the sheath in a manner that require relatively less characteristic length necessary to provide the shielding of potential given at the probe surface.

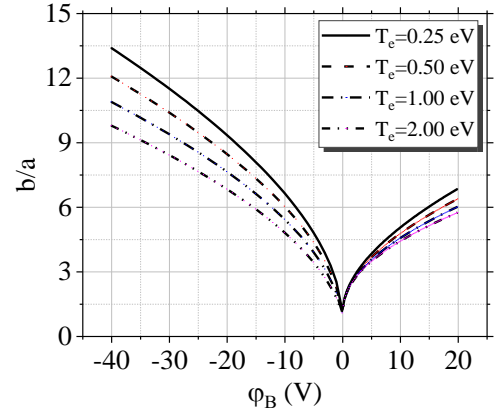

*Figure 1: Plot of normalised sheath radius ($b/a$) as a function of applied bias ($\varphi_B$) for varying electron temperature ($T_e$) at constant electron density, $n_e = 3 \times 10^{15}$.*

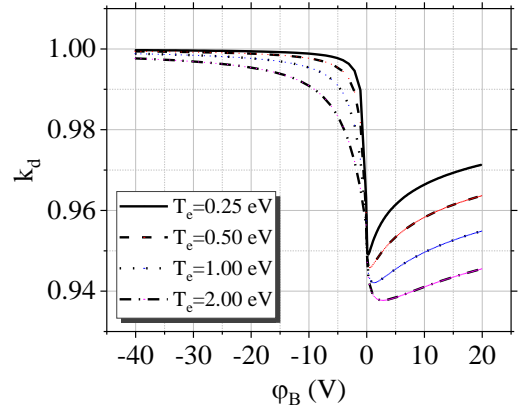

*Figure 2: Plot of sheath dielectric ($k_d$) with applied bias ($\varphi_B$) for varying electron temperature ($T_e$).*

Once the sheath radius ($b$) and sheath potential profile $\varphi(r)$ is determined, sheath dielectric ($k_d$) can also be evaluated using Eqs. (8) and (9). Fig. (2) shows the variation in sheath dielectric for different values of $T_e$. It is observed that for a large negative bias ($\varphi_B \ll 0$), sheath dielectric is approaching towards the vacuum value ($k_d \rightarrow 1$), signifying the fact that the sheath is almost electron-free. However, $k_d$ decreases close to the plasma potential ($V_p$) which is a direct function of average electron population [Eq. (7)] inside the sheath, $\bar{n}_e$. For a fixed applied negative bias, $k_d$ is observed to decrease further with the increase in $T_e$. This is due to the increase in $\bar{n}_e$ as a result of rise in electron mean energy.

On the other hand, in the electron sheath ($\varphi_B > 0$), $k_d$ continues to decline and then increases [Fig. (2)]. The minimum value in increasing and the rising slope is decreasing with the increase in $T_e$. This is due to the fact that following the continuity equation inside the cylindrical sheath, the electron density inside the sheath falls and then increases near the cylindrical probe surface, as generally seen in the ionic sheath [32,33]. Therefore, for small values of applied bias, the average electron density inside the sheath ($\bar{n}_e$), exceed the density at the bulk. Nonetheless, $\bar{n}_e$ decreases for further applied bias, as evident with the increase in $k_d$, with the rising slope being a function of $T_e$.

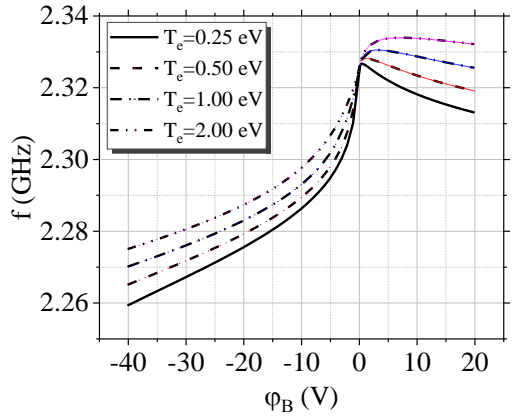

*Figure 3: Plot of resonance frequency ($f$) versus applied bias ($\varphi_B$) for varying electron temperature.*

The sheath radius $b$ [Fig. (1)] along with $k_d$ [Fig. (2)] can be used in Eq. (4) to obtain the resonance frequency ($f_r$) for an assumed value of vacuum resonance frequency, $f_0 = 2.235$ GHz. This is shown in Fig.-3. It is seen that $f_r$ increase with the applied bias, reaches a maximum and then fall afterwards. It is also observed that with the increase in $T_e$, the maximum value of $f_r$ shifts towards higher probe bias. Since $f_r$ is mainly characterized by $b$ (or $\Lambda(V_B)$) and $k_d$; the shift in the peak value of $f_r$ attributes directly to the shift in $k_d$.

## 4. Experimental Setup and Diagnostic:

To validate the effect of electron and ion sheath on the resonance frequency of a hairpin probe, the experiment is carried out in a capacitively coupled plasma (CCP) excited at a frequency of 13.56 MHz using argon as feedstock gas.

Fig. (4) shows the schematic of experimental setup. It consists of two annular parallel plate of radii $a = 5\ cm$ and $b = 10\ cm$, separated by a distance $d = 10\ cm$. The radio frequency power supply (RFG-600W Coaxial Power) is applied across the plates in push-pull configuration to create the plasma. The source assembly is placed inside a grounded vacuum chamber of $R = 15\ cm$ radius and $1.5\ m$ length. This source provides a uniform plasma density in the central hollow part, i.e., 20-mm radius [34].

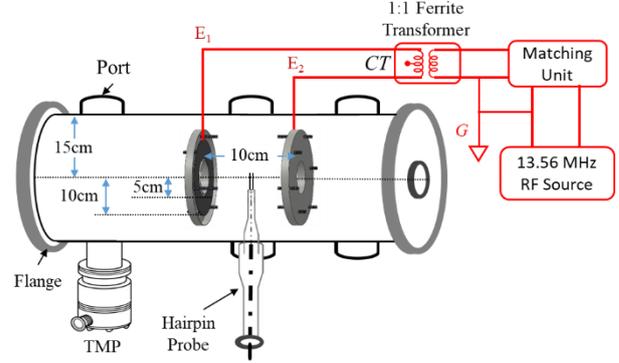

*Figure 4: Schematic of the experimental setup.*

The HP consists of a thin 'U' shape tungsten wire, which is inductively coupled with a loop antenna, as shown in Fig. (5). The loop antenna is formed by shorting the central conductor of a co-axial semi-rigid copper rod (RG-405) to its grounded shield and placed in the central hole of a multi-hole ceramic tube. The hairpin is supported along with the electrical connections through one of the side holes of the ceramic tube.

A voltage controlled YIG (Yttrium Iron Garnet) microwave oscillator (frequency range 1–18 GHz, power level 10 dBm) is used to provide a frequency sweep to the HP via passive directional coupler. At resonance, maximum power is transmitted due to a strong coupling between the loop antenna and the hairpin. A Schottky diode (PE 2210-10) converts this reflected frequency signal to inverted voltage signal. A LabVIEW programme controls the voltage steps applied to the oscillator using a data acquisition and control unit (National Instruments NIUSB 6251). The rms value of the reflected microwave signal is then obtained on a Tektronix 3034C digital storage oscilloscope and sent to a computer using a GPIB-USB converter.

A linear amplifier (WMA-300, Falco Systems) is used to amplify a sweep from a signal generator (Tektronix-AGF3021C) and provided to the probe for as a DC bias. The corresponding change in the reflected signal for each frequency step is displayed on the oscilloscope (Tektronix 3054C) and acquired using LabVIEW software.

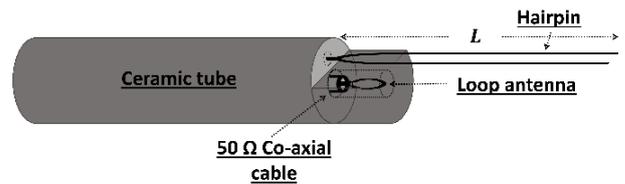

*Figure 5: Configuration of Hairpin probe.*

The vacuum resonance frequency $f_0$ of HP has been found to be 2.28 GHz with a quality factor of 109.5 for the resonance frequency peak. This agrees well with the theoretically obtained value $f_0 = c/2(2L + w) = 2.29$ GHz corresponding the length 31.5 mm and width 2.5 mm.

## 5. Experimental results and analysis:

The experiment is carried out in argon discharge at a constant pressure and power of 7.5 mTorr and 50 Watt, respectively. Fig. (6) show the resonance frequency as a function of applied bias. The resonance frequency ($f_r$) is observed to be monotonically increasing attaining a maximum value followed by a steady fall at higher probe bias.

The hairpin probe utilized in the investigation is about one-third the length of the central hollow part of the discharge. When the hairpin probe is near the electron saturation current, it could disturb the adjacent plasma while collecting electrons. To confirm this, the hairpin probe's maximum applied bias is adjusted, as seen in Fig. 6. It is found that the resonance frequency characteristic retraces itself [see Fig. 6(a)–6(f)], demonstrating that the findings are not an artefact of local plasma disturbance. Instead of a sharp peak, as reported previously in various articles [6,23,35], a steady fall with a constant slope is observed endorsing the significant impact of sheath dielectric, as discussed in Sec. (2).

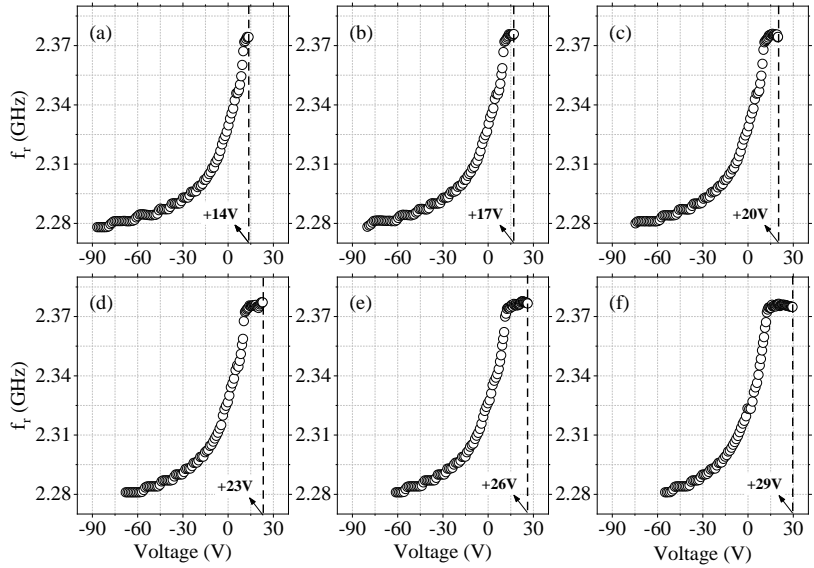

Figure 6: Plot of resonance frequency ($f$) versus applied bias ($V_B$) to HP with the increase in the positive biased voltages.

As discussed in section 2, the resonance frequency characteristic is correlated to the change in the plasma parameters such as $V_P$, $n_e$ and $T_e$. By varying $V_P$, the resonance frequency curve shifts horizontally [Fig. 7], while changes in $n_e$ induce a vertical shift [Fig. 8]. As depicted in Fig. 9, the observed peak in the experimentally obtained characteristic curve is utilized to derive an initial estimate for $V_P$ and $n_e$ using Eq. (3a) where $\xi = 1$, i.e., $n_e(10^{16}m^{-3}) = (f_r^2 - f_0^2)/0.81$.

With these initial estimations of $V_P$ and $n_e$, the electron temperature ($T_e$) is iteratively varied to achieve a close fit with the experimental curve. Eventually, all three parameters undergo variations to determine the best fit, as shown in Fig. (10). The estimated values of $T_e$, $n_e$, and $V_P$, respectively, are $2.3\ eV$, $5.3 \times 10^{15}\ m^{-3}$, and 10.6 V. In comparison to these values, those determined using the hairpin probe assisted compensated Langmuir probe [36] are $2.2\ eV$, $6.17 \times 10^{15}\ m^{-3}$, and 11.4 V, i.e., the model fit parameters are within 10–15 percent error of the Langmuir probe results.

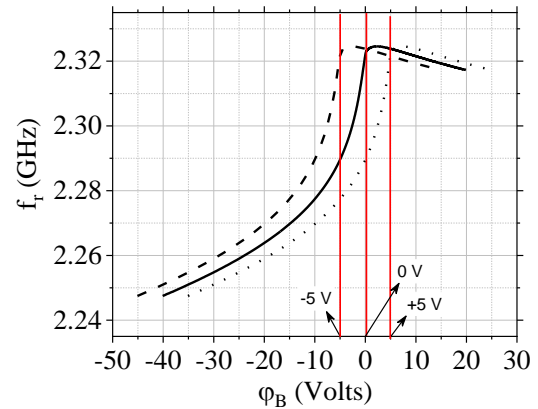

Figure 7: Plot represents a horizontal shift in the resonance characteristic curve with the change in plasma potential.

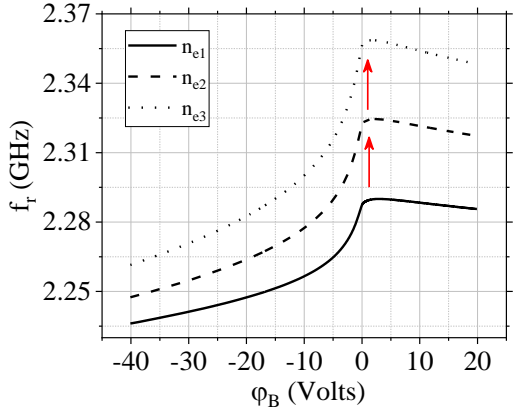

*Figure 8: Plot represents a vertical shift in the resonance characteristic curve with the increase in the electron density. Here, $n_{e1}$, $n_{e2}$, and $n_{e3}$ are respectively 3, 5, and $7 \times 10^{15} m^{-3}$.*

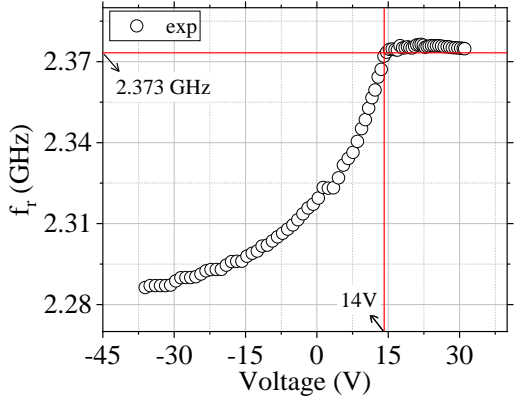

*Figure 9: Plot representing the initial guess values of plasma potential ($V_P = +14\,V$) and electron density $[n_e(10^{16}m^{-3}) = (f_r^2 - f_0^2)/0.81]$ corresponding to the resonance frequency ($f_r = 2.373\,GHz$).*

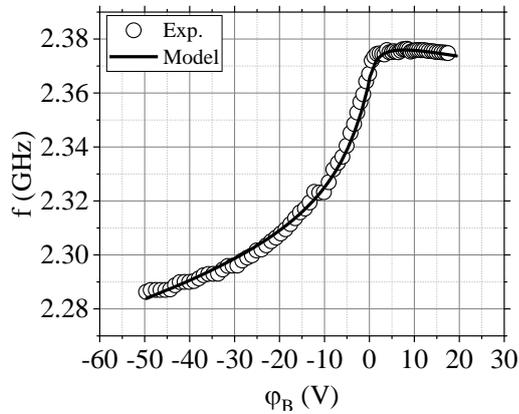

*Figure 10: Plot of experimental and theoretical generated values of resonance frequency ($f$) as a function of applied bias ($\varphi_B$) in argon plasma.*

## 6. Discussion and Conclusion:

The plasma parameters extracted from the present model and Haas's model [37] can be compared to illustrate the improvements made over Haas's model. However, it is observed that achieving a best match that covers the complete range, including both the ion and electron sheath parts, is not feasible. This limitation is illustrated in Figs. 11 and 12, where the former models the ionic part only, and the latter models the electronic part exclusively.

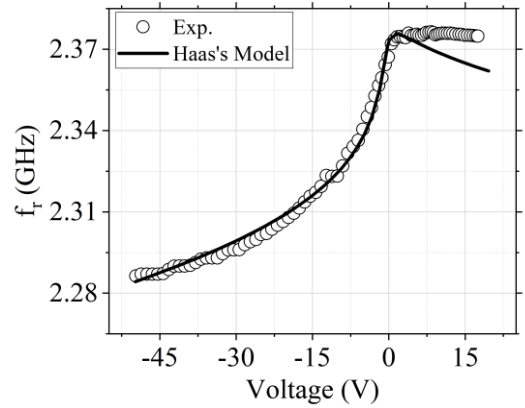

*Figure 11: Plot represents the fitting of experimental data with Haas's model for the ion sheath region ($\varphi_B < 0$).*

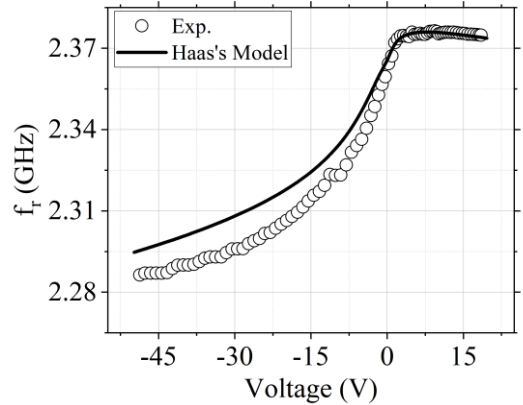

*Figure 12: Plot represents the fitting of experimental data with Haas's model for the electron sheath region ($\varphi_B > 0$).*

In the case of an electron sheath, the electron density at the sheath edge is taken as bulk electron density, whereas the velocity is its respective thermal velocity. This would be a reasonable assumption given that the potential difference across the electron pre-sheath is significantly smaller than that across the electron sheath. However, some recent theoretical and numerical studies [28–30,38,39] have revealed the possibility of the existence of electron Bohm sheath criteria; hence, electron presheath. Electrons in the presheath are found to acquire large drift velocity even in small potential rise of the order of positive ion temperature ($T_+$). This is due to the dominating pressure term in the electron momentum equation. On the other hand, positive ions will be following the Boltzmann distribution in the presheath. This gives rise to an overall plasma density fall in the presheath region similar to that of ionic presheath as $n_{+s} = n_+ exp(e\varphi_{presheath}/T_+) \sim n_+ exp(-1) \sim 0.398 n_+$. The proposed Bohm velocity equivalence of electrons is $v_{eB} = \sqrt{eT_e/m_e}\sqrt{1 + 1/\gamma_+}$ [28]. Here, $\gamma_+$ is the electron-to-positive ion temperature ratio.

Fig.-11 presents the $f_r$ vs $V_B$ curves when the electron pre-sheath is incorporated for varying electron temperature. It is found that the resonance frequency always acquires a peak at the plasma potential, which is in contrast to the case when electron presheath is excluded [Fig. 3]. However, the experimental curve [Fig. 6] suggests that the resonance frequency curve somewhat saturates at positive applied bias with respect to the plasma potential. This circumstantial evidence suggests the absence of electron pre-sheath, similar to the claim of [40].

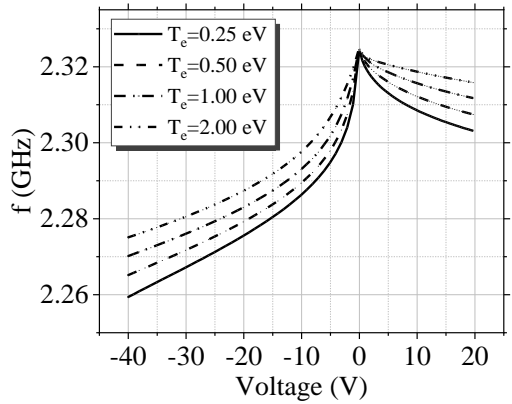

*Figure 13: Plot of resonance frequency versus applied bias ($\varphi_B$) for varying electron temperature. Here, $n_e = 3 \times 10^{15}$.*

Piejak et al. [24] have shown that the plasma potential can be determined from the peak of the resonance frequency characteristic curve. Later, this is also validated in a DC discharge by comparing the peak value with the first derivative of the current-voltage characteristic of a cylindrical Langmuir probe [16]. However, in the present study the resonance frequency characteristic exhibits a broadening near the maxima in the resonance frequency curve, instead of a sharp peak. This makes it further difficult for deducing the plasma potential using this method. Nonetheless, for low electron temperature plasma ($T_e < 0.5\ eV$), the peak in the resonance frequency can still be reasonably used as a measure of plasma potential ($V_P$).

Under the present analysis, both ion and electron sheaths are under collision-less regime. Furthermore, the present work assumes the background positive ions to be cold such that $\beta = T_+/T_e \to 0$. Under these conditions, the ions are expected to follow a quasi-radial trajectory when falling toward the probe [41]. However, the simple presumption of radial electron motion would underestimate the negative space charge that might exist if orbital trajectories had been included. This refinement to the model could be further explored.

It is also important to emphasize that the ceramic casing of the loop antenna [Fig. 5] would also form an ion sheath around it, which could interfere with the sheath formed around the hairpin itself. However, the short-circuit end of the hairpin is less sensitive than the probe pins. In this particular case, the dielectric covering the loop is less than 10% of the total length; therefore, it will have a minimal effect on the resonance frequency obtained. Alternatively, if the dielectric is greater than 30%, then one has to apply a correction model [42].

In summary, this work presents the effect of thermal electrons on the resonance frequency of a DC-biased hairpin probe. It has been found that the resonance frequency characteristic curve is primarily a function of electron temperature. The plasma potential and electron density merely shift the curve in a horizontal and vertical direction, respectively. With the increase in electron temperature, the maximum value of a characteristic curve shows a broadening. This broadening limits the estimation of plasma potential to low-temperature plasmas. Furthermore, by comparing the resonance frequency curve obtained using this model with the one obtained in the experiment, an attempt is made to estimate the plasma parameters, such as plasma potential, absolute electron density, and electron temperature, in a capacitively coupled RF argon discharge. Since, the work assumes cold plasma approximation for plasma dielectric However, the applicability of the present measurement methodology to a low-pressure condition such that collisions do not play a significant role, i.e., pressures<< 100 mTorr [26,41].

## Acknowledgment

This work is supported by the Department of Atomic Energy, Government of India. Dr. N. Sirse is supported by the Science and Engineering Research Board (SERB) Core Research Grant No. CRG/2021/003536.